\documentclass[11pt]{article}
\usepackage{amsmath,amsfonts,amsthm}

\newtheorem{Def}{Definition}
\newtheorem{The}{Theorem}
\newtheorem{Pro}{Proposition}

\theoremstyle{definition}
\newtheorem{Rem}{Remark}

\numberwithin{equation}{section}

\begin{document}
\title{On Classification of Integrable Davey-Stewartson Type Equations}
\author{Benoit Huard$^{\dagger}$, Vladimir Novikov$^{*}$}

\maketitle
\begin{center}
$^{\dagger}$ Department of Mathematics, Northumbria University, \\
Newcastle upon Tyne, United Kingdom\\[1cm]
$^*$ Department of Mathematical Sciences, Loughborough University, \\Loughborough, Leicestershire LE11 3TU \\ United Kingdom \\[2ex]
e-mail: \\[1ex]
\texttt{Benoit.Huard@northumbria.ac.uk, V.Novikov@lboro.ac.uk}
\end{center}

\begin{abstract}
This paper is devoted to the classification of integrable Davey-Stewartson type equations. A list of potentially deformable dispersionless systems is obtained through the requirement that such systems must be generated by a polynomial dispersionless Lax pair. A perturbative approach based on the method of hydrodynamic reductions is employed to recover the integrable systems along with their Lax pairs. Some of the found systems seem to be new.\\

\noindent {MSC:} 35L40, 35Q51, 35Q55, 37K10.\\
\noindent {PACS:} 02.30.Ik, 02.30.Jr\\[5mm]
\noindent {Keywords:} Davey-Stewartson equations, Dispersive deformations, Hydrodynamic reductions, Lax pairs.
\end{abstract}

\section{Introduction}

In this paper we continue the programme of classification of
integrable dispersive $2+1$-dimensional equations started in
\cite{FMN,FN} and present the complete classification of integrable
 two-component systems of second order
\begin{equation}
\label{sysgen}
\begin{split}
&u_t =F(u,v,w,Du,Dv,Dw) \\
&v_t = G(u,v,w,Du,Dv,Dw)
\end{split},
 \quad w_y = u_x.
\end{equation}
Here $u(x,y,t), v(x,y,t), w(x,y,t)$ are scalar variables, $Du,Dv,Dw$ denote the collection of partial derivatives of $u,v,w$ with respect
to $x,y$ up to the second order and $F$, $G$ are polynomials in derivatives with coefficient depending only on $u,v,w$.

The approach proposed in \cite{FMN,FN} consists of two steps:
\begin{itemize}
\item One first classifies integrable {\it dispersionless}
equations which may potentially occur as dispersionless limits of
systems in consideration;
\item One then reconstructs the {\it dispersive} corrections
preserving integrability.
\end{itemize}

One of the most famous examples within class (\ref{sysgen}) was introduced by Zakharov in \cite{Zakharov}
\begin{equation}
\label{dispersive-Zakharov}
u_t = (uv)_x + u_{xx}, \quad
v_t = v v_x + w_x - v_{xx}, \quad
w_y = u_x,
\end{equation}
together with the associated Lax pair
\begin{equation}
\label{dispersive-Zakharov-Lax}
\psi_{xy}+\frac{1}{2} v\psi_y+\frac{1}{4}u\psi=0, \quad \psi_t+ \psi_{xx}+\frac{1}{2}w\psi=0.
\end{equation}
Two more known examples \cite{MikYam1997,ShabYam1997} include
\begin{equation*}
u_t=wu_x+v_x+u_{xx},\quad v_t=(vw)_x-v_{xx},\quad w_y=u_x
\end{equation*}
and
\begin{equation*}
u_t=(uv)_x+wu_x+u_{xx},\quad v_t=(vw)_x+vv_x-v_{xx}, \quad w_y=u_x
\end{equation*}
and their Lax representations are given by
$$
4\psi_{xy}+v\psi-2 u\psi_x=0,\quad \psi_t+\psi_{xx}-w\psi_x=0
$$
and
$$
2\psi_{xy}+v\psi_y-u\psi_x=0,\quad \psi_t+\psi_{xx}-w\psi_x=0.
$$
The classification results obtained in this paper include 6 more integrable systems of the form (\ref{sysgen}) (Theorem 3), some of which seem to be new.

The dispersionless counterpart of system
(\ref{dispersive-Zakharov}) read as
\begin{equation}
\label{dZakin}
u_t = (uv)_x, \quad
v_t = v v_x + w_x, \quad
w_y =
u_x
\end{equation}
and the corresponding Lax representation is
\begin{equation}
\label{displess-Zakharov-Lax}
S_xS_y+\frac{1}{2}vS_y+\frac{1}{4}u=0,\quad S_t+S_x^2+\frac{1}{2}w=0.
\end{equation}
These can be obtained from (\ref{dispersive-Zakharov}) and
(\ref{dispersive-Zakharov-Lax}) through a change
\begin{eqnarray}
\label{displim} x\to \epsilon x,\, y\to\epsilon y,\, t\to\epsilon
t,\quad \psi=e^{\frac{S}{\epsilon}}
\end{eqnarray}
and taking the limit $\epsilon\to 0$.


We recall that the dispersionless system (\ref{dZakin}) is
integrable by the method of hydrodynamic reductions (see e.g.
\cite{FerKhu}), a method which can be applied to any first-order
quasilinear system
\begin{equation}
\label{displess-system}
A(u) u_t + B(u) u_x + C(u) u_y = 0,
\end{equation}
where $A, B, C$ are $n\times n$ square matrices and $u$ is an $n$-dimensional vector. Integrability in this sense means that for any fixed number $N$, an integrable system of the form (\ref{displess-system}) possesses an infinite class of solutions of the form
\begin{equation}
\label{displess-solution}
u=u(R^1, \ldots, R^N), \quad v=v(R^1, \ldots, R^N), \quad w=w(R^1, \ldots, R^N),
\end{equation}
parametrized by $N$ arbitrary functions of a single argument, where the Riemann invariants $R^i$ (also called phases) are assumed to satisfy pairs of commuting diagonal systems
\begin{equation}
\label{Riemann-invariants}
R^i_y = \mu^i(R) R^i_x, \quad R^i_t = \lambda^i(R) R^i_x, \quad i=1,\ldots,N.
\end{equation}
In order that relations (\ref{displess-solution}) and (\ref{Riemann-invariants}) represent a nontrivial solution, it is readily seen that the velocities $\lambda^i, \mu^i$ must satisfy the so-called dispersion relation
\begin{equation}
\label{dispersion-relation}
D(\lambda^i,\mu^i) = \det{\left(\lambda^i A + B + \mu^i C\right)} = 0, \quad i=1,\ldots,N.
\end{equation}
Throughout this paper, we consider only systems for which
(\ref{dispersion-relation}) defines an irreducible curve. Moreover,
compatibility of systems (\ref{Riemann-invariants}) requires that
the velocities $\lambda^i(R), \mu^i(R)$ obey the conditions
\begin{equation}
\label{comp-diag}
\frac{\lambda^i_{j}}{\lambda^i - \lambda^j} = \frac{\mu^i_{j}}{\mu^i - \mu^j}, \quad i\neq j,
\end{equation}
where  $\lambda^i_j = \partial_{R^j}\lambda^i, \mu^i_j = \partial_{R^j}\mu^i$.  We admit the following definition of integrability of dispersionless systems

\begin{Def}
A quasilinear system  (\ref{displess-system}) is said to be integrable if for any number of
phases $N$ it possesses infinitely many $N$-phase solutions
parametrised by $N$ arbitrary functions of one variable.
\end{Def}
Owing to the fact that the flows (\ref{Riemann-invariants}) are automatically commuting for $N=1$, one-component reductions of system (\ref{dZakin}) are parametrized by the following relations
\begin{equation}
\label{one-component-dZak}
\begin{split}
&u = R, \quad v = V(R), \quad w = W(R), \quad 
R_t = (R V' + V) R_x, \quad R_y = \frac{1}{R (V')^2} R_x,
\end{split}
\end{equation}
where $V(R)$ is an arbitrary function and $W(R)$ is fixed up to an
integration constant by the relation $W' = R (V')^2$. Moreover, the
normalization $u=R$ is assumed to hold through a Miura-type
transformation. It was shown in \cite{FMN,FN,FM} that deformability
of hydrodynamic reductions can be used as a definition of
integrability of dispersive equations. Indeed, formulae
(\ref{displess-solution}), (\ref{Riemann-invariants}) governing
hydrodynamic reductions can be deformed through the addition of a
dispersive correction of the form
\begin{equation}
\label{eq:dispersive-corrections-u}
\begin{split}
&u=u(R^1,...,R^n), \quad v=V(R^1,...,R^n) + \epsilon (\ldots) + \epsilon^2 (\ldots) + \ldots, \\
&w=W(R^1,...,R^n) + \epsilon (\ldots) + \epsilon^2 (\ldots) + \ldots,\\
&R^i_t = \lambda^i(R) R^i_x + \epsilon (\ldots) + \epsilon^2 (\ldots)+ \ldots , \quad R^i_y = \mu^i(R) R^i_x + \epsilon (\ldots) + \epsilon^2 (\ldots) + \ldots,
\end{split}
\end{equation}
where the terms at $\epsilon^k$ are homogeneous differential polynomials in the $x$-derivatives of $R^i$ of total degree $k+1$ with coefficients being functions of the Riemann invariants.  Using this construction the following definition of integrability of $2+1$-dimensional dispersive systems was introduced in \cite{FMN}.

\begin{Def} A $2+1$-dimensional system is said to be integrable if
all hydrodynamic reductions of its dispersionless limit\footnote{which is
supposed to be linearly non-degenerate, see Remark \ref{linear-nondegeneracy}.} can be deformed into
reductions of the corresponding dispersive counterpart.
\end{Def}

To accomplish the classification program, the polynomial systems (\ref{sysgen}) are written in the quasilinear form
\begin{equation}
\label{DS}
\begin{split}
&u_t = \alpha u_x + \beta u_y + \gamma v_x + \delta v_y + \rho w_x + \epsilon( \ldots )\\
&v_t = \phi u_x + \psi u_y + \eta v_x + \tau v_y + \kappa w_x + \epsilon( \ldots )
\end{split}
, \quad
w_y = u_x,
\end{equation}
under which they are often referred to as Davey-Stewartson type systems.
Here the greek symbols $\alpha, \ldots,$ denote functions of $u,v$ and the nonlocal variable $w = D_y^{-1} D_x u$
and the terms at $\epsilon$ are assumed to be homogeneous differential polynomials of total degree $2$. The dispersionless limit
 \begin{equation}
\label{dDS}
\begin{split}
&u_t = \alpha u_x + \beta u_y + \gamma v_x + \delta v_y + \rho w_x \\
&v_t = \phi u_x + \psi u_y + \eta v_x + \tau v_y + \kappa w_x
\end{split}
, \quad
w_y = u_x,
\end{equation}
is thus obtained by simply setting $\epsilon=0$. To classify integrable systems (\ref{DS}) we first have to classify integrable {\it dispersionless} systems (\ref{dDS}) and then classify dispersive deformations of the obtained systems. Classification of integrable systems (\ref{dDS}) can be performed using the conditions of existence of 2- and 3-component hydrodynamic reductions (Section 2). The list of such systems is very extensive and moreover most of examples are not deformable to (\ref{DS}). However we notice  that if a {\it dispersive} system (\ref{DS}) is
integrable then it possesses a Lax representation
$$
L\psi=0,\quad (\partial_t-A)\psi=0,
$$
where $L$ and $A$ are some linear differential operators with
coefficients being functions in $u,v,w$ and their derivatives. The
dispersionless limit (\ref{displim}) brings the Lax representation
to a form
$$
F(S_x,S_y)=0,\quad S_t=G(S_x,S_y),
$$
where $F$ and $G$ are {\it polynomials} in $S_x,S_y$ with
coefficients depending on $u,v,w$. Thus if a dispersionless system
is deformable then it possesses a {\it polynomial} dispersionless
Lax representation and it is enough to restrict ourself to only such dispersionless systems. We therefore adopt the following classification approach:
\begin{itemize}
\item We first classify integrable dispersionless systems which possess polynomial dispersionless Lax representations (Section 2). In this paper we restrict ourselves to second degree polynomials;
\item We then classify dispersive deformations of such systems (Section 3).
\end{itemize}

We finally emphasize that the same perturbative approach can be used to
obtain the Lax pair of an integrable {\it dispersive} system from
that of its {\it dispersionless} limit, given that this limit is integrable in the sense of hydrodynamic reductions. In Section 3 we present a construction which allows to deform dispersionless Lax pairs into their dispersive counterparts and give the complete list of integrable systems (\ref{DS}) along with associated Lax representations.



\section{Classification of dispersionless systems.}
\label{sec:DispLaxPairs}

In this section we describe and apply the two approaches to the classification of integrable dispersionless systems
\begin{equation}
\label{disples-sys} {\cal E:} \left\{ \begin{array}{l}
u_t=\alpha u_x+\beta u_y+\gamma v_x+\delta v_y+\rho w_x\\
v_t=\phi u_x+\psi u_y+\eta v_x+\tau v_y+\kappa w_x\end{array},\quad
w_y=u_x, \right.
\end{equation}
where we assume that coefficients $\alpha,\ldots,\kappa$ are functions of $u,v,w$.

On one hand, the method of hydrodynamic reductions provides a sufficiently restrictive integrability criterion which permits to distinguish integrable systems within the considered class. Similarly to the 3-soliton criterion for solitonic systems, it can be seen that the presence of 3-phase solutions  (solutions of the form (\ref{displess-solution}), (\ref{Riemann-invariants}) with $N=3$) is necessary and sufficient to prove integrability of dispersionless systems. However, requiring directly the existence of $3$-phase solutions for system (\ref{disples-sys}) is hardly workable since the dispersion relation of system (\ref{disples-sys}) defines a cubic curve which may not be {\it a priori} a rational curve. The following theorem allows to overcome this difficulty by providing necessary and sufficient conditions for the existence of 2-phase solutions for any system (\ref{displess-system}).
\begin{The}[Ferapontov, Khusnutdinova \cite{FK}]
The Haantjes tensor of an arbitrary matrix $\Omega = \left(k A + B\right)^{-1} \left(l A + C\right)$, $k,l \in \mathbb{R}$, is zero if and only if the following hold.
\begin{itemize}
\item System (\ref{displess-system}) possesses double waves parametrized by four arbitrary
functions of a single argument.
\item The characteristic speeds $\lambda^i$, $\mu^i$ of two-component reductions are not
restricted by any algebraic relations other than the dispersion relation (\ref{dispersion-relation}).
\end{itemize}
\end{The}
Recall that for a given square matrix $\Omega=(\omega^i_j)$, the Haantjes tensor is defined as \cite{Haantjes}
\begin{equation}
H^i_{jk} = N^i_{pr} \omega^p_j \omega^r_k - N^p_{jr} \omega^i_p \omega^r_k - N^p_{rk} \omega^i_p \omega^r_j + N^p_{jk} \omega^i_r \omega^r_p,
\end{equation}
where the $N^i_{jk}$ are the components of the Nijenhuis tensor
\begin{equation}
N^i_{jk} = \omega^p_j \partial_{u^p} \omega^i_k - \omega^p_k \partial_{u^p} \omega^i_j - \omega^i_p (\partial_{u^j} \omega^p_k - \partial_{u^k} \omega^p_j).
\end{equation}
Thus the vanishing of the Haantjes tensor imposes a large set of first order differential constraints on the ten unknown functions $\alpha, \ldots, \kappa$. These conditions are very restrictive and allow to obtain a wide list of systems (\ref{dDS}) which possess 2-component reductions. Imposing the further condition of existence of 3-component reductions one then obtains the complete list of integrable dispersionless systems (\ref{dDS}). The list is extremely lengthy and most of systems are not deformable to (\ref{DS}).


\begin{Rem}
More precisely, it was possible to solve completely the case $\rho = 0$, while the case $\rho \neq 0$ seemed to be out of reach for the moment. Nevertheless, it is remarkable that all {\it deformable} integrable systems obtained through our second approach fall into the class $\rho = 0$.
\end{Rem}

\begin{Rem}
\label{linear-nondegeneracy}
In order to apply the perturbative approach it is necessary to exclude linearly degenerate dispersionless systems as their dispersive counterparts do not inherit the hydrodynamic reductions. These systems are characterized by the fact that their $N$-phase solutions are continuous and do not admit the gradient catastrophe. A different perturbative approach would be required. Fortunately, these systems can be easily discarded by verifying an algebraic criterion that was introduced in \cite{F1}.
\end{Rem}


Now we proceed to the classification of dispersionless systems (\ref{disples-sys}) which possess polynomial dispersionless Lax representations. It was noted in the Introduction that the system
\begin{equation}
\label{dzak1}
u_t = (uv)_x, \quad
v_t = v v_x + w_x, \quad
w_y =
u_x,
\end{equation}
possesses a dispersionless Lax representation
\begin{equation}
\label{zakdlp}
S_xS_y+\frac{1}{2}vS_y+\frac{1}{4}u=0,\quad S_t+S_x^2+\frac{1}{2}w=0.
\end{equation}
Introducing notations $p=S_x,\,q=S_y,\,s=S_t$ the relations (\ref{zakdlp}) can be rewritten as
\begin{equation}
\label{zakdlp1}
F:=pq+\frac{1}{2}vq+\frac{1}{4}u=0,\quad s=G:=-p^2-\frac{1}{2}w
\end{equation}
and the compatibility conditions $S_{xy}=S_{yx},S_{xt}=S_{tx},S_{yt}=S_{ty}$ take the form
\begin{equation}
\label{comp}
p_y=q_x,\quad p_t=s_x,\quad q_t=s_y.
\end{equation}
It is easy to see that relations (\ref{zakdlp1}) and (\ref{comp}) can be written as
\begin{equation}
\label{zakdlp2}
F_t=\{G,F\}\quad\mod F,\mbox{equations (\ref{dzak1})},
\end{equation}
where $$\{G,F\}:=\frac{\partial G}{\partial p}\frac{\partial
F}{\partial x}-\frac{\partial F}{\partial p}\frac{\partial
G}{\partial x}+\frac{\partial G}{\partial q}\frac{\partial
F}{\partial y}-\frac{\partial F}{\partial q}\frac{\partial
G}{\partial y}.$$
and $\frac{\partial F}{\partial x}=u_x\frac{\partial F}{\partial u}+v_x\frac{\partial F}{\partial v},\ldots$, i.e $p,q$ are treated as symbols in (\ref{zakdlp2}).

We now introduce the following definition of polynomial dispersionless Lax pair (see eg \cite{Zakharov, KonMag}).

\begin{Def} A pair $F(p,q), G(p,q)$ of polynomials in $p,q$ with
coefficients depending on $u,v$ and $u,v,w$ correspondingly is
called a generator of a dispersionless Lax pair for a system
(\ref{disples-sys}) if
\begin{enumerate}
\item $F,G$ satisfy
\begin{equation}
\label{Lax} F_t=\{G,F\}\quad \mod F,{\cal E}
\end{equation}

\item $\mbox{rank}\left(\frac{\partial F_{ij}}{\partial u}\,\,\frac{\partial F_{ij}}{\partial
v}\right)_{i,j=1,\ldots,\deg{F}}=2$, where
$F_{ij}=\frac{\partial^{i+j} F}{\partial p^i\partial
q^j}\big|_{p=q=0}$.

\end{enumerate}
\end{Def}

A polynomial Lax pair for a given system is not unique. Apart from transformations $p\to \mu_1 p+\mu_2,\, q\to \mu_1 q+\mu_3,\, s\to \mu_1 s+\mu_4$, $\mu_1,\ldots\mu_4=const$ we also have that if $F, G$ are generators of a dispersionless Lax pair for a system then $f(F), G$  are also generators of a dispersionless  Lax pair for the same system for an arbitrary polynomial $f$. Moreover $G$ is defined modulo $F$. Condition 1 is the compatibility condition while condition 2 implies that the Lax pair is faithful, i.e.  the compatibility condition is satisfied due to both equations in the system (\ref{disples-sys}).

We now present the list of dispersionless systems which possess polynomial dispersionless Lax pairs. We restrict the classification of polynomial Lax pairs to those of degree 2 and leave the consideration of systems with Lax pairs of higher degrees for future studies.

\begin{The} If the system (\ref{disples-sys}) possesses a polynomial dispersionless Lax pair with $\deg{F}=2,\,\,\deg{G}\mod F=2$ and if $F$ is an irreducible polynomial then up to the group of invertible transformations $v\to f(u,v)$,
 $x\to x,\,y\to y+sx,\, u \to u, w\to w-s u,\, s=const$,
and rescaling of variables it is one of the following ones:

\begin{eqnarray}
\label{dZak} &&\left\{\begin{array}{l} u_t=(uv)_x\\ v_t=vv_x+w_x\end{array},\right.
\\ \label{de1} &&\left\{\begin{array}{l} u_t=wu_x+v_x\\ v_t=(vw)_x\end{array},\right.
\\ \label{de2} &&\left\{\begin{array}{l} u_t=wu_x+(uv)_x\\ v_t=(vw)_x+vv_x\end{array},\right.
\\ \label{de3} &&\left\{\begin{array}{l} u_t=wu_x+v_x-\frac{h^2}{4}vv_y\\ v_t=(vw)_x+hvv_x\end{array},\right.
\\ \label{de4} &&\left\{\begin{array}{l} u_t=uu_x-wu_y+v_x+vv_y\\ v_t=2vu_x+v^2u_y+uv_x-wu_y+w_x\end{array},\right.
\\ \label{de5} &&\left\{\begin{array}{l} u_t=(v-hu)u_x+h(w-\frac{1}{2}v^2)u_y+uv_x-huvv_y\\ v_t=-2hvu_x+h^2v^2u_y+(v-hu)v_x+h(w-\frac{1}{2}v^2)v_y+w_x\end{array},\right.
\\ \label{de6} &&\left\{\begin{array}{l} u_t=(hu+v)u_x+h(w-2huv-\frac{1}{2}v^2)u_y+uv_x-hu(hu+v)v_y\\ v_t=-2hvu_x+h^2v^2u_y+(v-hu)v_x+h(w-\frac{1}{2}v^2)v_y+w_x\end{array},\right.
\\ \label{de7} &&\left\{\begin{array}{l} u_t=(uv)_x+wu_x-\frac{h}{2}v^2(1+2hu)u_y-huv(1+hu)v_y\\ v_t=(vw)_x+vv_x-2hv^2u_x+h^2v^3u_y-\frac{1}{2}hv^2(1-2hu)v_y-2huvv_x
\end{array},\right.
\\ \label{de8} &&\left\{\begin{array}{l} u_t=(w-2huv)u_x+h^2uv^2u_y+(1-hu^2)v_x+hv(u^2-1)v_y\\ v_t=-2hv^2u_x+h^2v^3u_y+(w-2huv)v_x+h^2uv^2v_y+vw_x
\end{array},\right.
\end{eqnarray}
where $w_y=u_x$, $h=const$.
\end{The}

By the $\deg{G}\mod F$ we understand the minimal degree of $G$ modulo $F$. Parameter $h$ in the above list, if not zero, can be scaled to $1$. Note that if $h\to 0$ then systems (\ref{de5}), (\ref{de6}) reduce to (\ref{dZak}), while systems (\ref{de7}), (\ref{de8}) reduce to (\ref{de2}) and (\ref{de1}) correspondingly. The corresponding generators for the  dispersionless Lax pairs are given in Table \ref{Table-displess-lax-pairs}.

\begin{table}[h]
\begin{math}
\begin{array}{|c|l|l|}
\hline
\text{Eq no} & $F$ & G \\\hline

(\ref{dZak}) & pq+\frac{1}{2}vq+\frac{1}{4}u & -p^2-\frac{1}{2}w \\[1ex] 

(\ref{de1}) & pq-\frac{1}{2}up+\frac{1}{4}v & -p^2+wp \\[1ex] 

(\ref{de2}) & pq-\frac{1}{2}u p+\frac{1}{2}vq & -p^2+w p \\[1ex] 

(\ref{de3}) & pq+\frac{1}{4}h v q+\frac{1}{4}v-\frac{1}{2}up-\frac{1}{8}huv & -p^2+wp-\frac{1}{16}h^2v^2 \\[1ex] 

(\ref{de4}) & pq+\frac{1}{2}u q+\frac{1}{4}+vq^2 & -p^2+v^2q^2-w q \\[1ex]

(\ref{de5}) & hpq-\frac{h}{2}(v+hu)q-h^2vq^2+\frac{1}{2}p & -p^2+h^2v^2q^2+\frac{1}{2}h(2w+v^2)q \\[1ex]

(\ref{de6}) & pq+\frac{1}{2}vq-hvq^2+\frac{1}{4}u  & -p^2+h^2v^2q^2-(h^2uv-hw+\frac{1}{2}hv^2)p-\frac{1}{2}w \\[1ex]

(\ref{de7}) & p q+\frac{p}{2h}(1-hu) - h v q^2+\frac{uv}{4} - \frac{1}{2} v q(1-h u)  & - p^2 + h^2 q^2 v^2 - \frac{1}{2}h u v^2 - \frac{1}{2} h q(2 h u-1) v^2 + w p \\[1ex]

(\ref{de8}) & pq - h v q^2 - \frac{1}{2}u p+\frac{v}{4} & -p^2 + h^2 v^2 q^2 + w p - \frac{1}{4} h v^2 \\ \hline
\end{array}
\end{math}
\caption{Generators for the Lax pairs of dispersionless systems (\ref{dZak})-(\ref{de8}).}
\label{Table-displess-lax-pairs}
\end{table}

We finally make the following observation which  greatly simplifies the computations entering the perturbation scheme in the next section.
\begin{Pro}
The dispersion relation (\ref{dispersion-relation}) of every system in (\ref{dZak})-(\ref{de8}) is a rational cubic curve.
\end{Pro}
Indeed the dispersion relation (\ref{dispersion-relation}) of systems of the form (\ref{disples-sys}) defines a homogeneous cubic polynomial in the variables $\lambda^i,\mu^i$ and can be viewed as a ternary cubic form by introducing triples $X^i, Y^i,Z^i$ through the relations $\lambda^i = X^i/Z^i, \mu^i = Y^i/Z^i$. It can be shown that for each of systems (\ref{dZak}) - (\ref{de8}) the dispersion relation is a singular cubic form and therefore admits a rational parametrization. This can be verified by noting that the discriminant of the curve vanishes identically. This discriminant takes the form of a sixth-order polynomial and there exists a straightforward algorithmic procedure originally developed by Hesse \cite{Hesse} to obtain its explicit expression  (see also the appendix in \cite{Nowak}). The singular point can either be a double point or a cusp and it should be noted that for systems (\ref{de5})-(\ref{de8}), the nature of the singularity can change in the limit $h \to 0$. For these cases, taking the limit in the corresponding Lax pair generators does not lead to a Lax pair of the limiting system.

\section{Dispersive deformations. The classification theorem.}
\label{sec:Deformations}

In this section we study the dispersive deformations of systems (\ref{dZak})-(\ref{de8}) up to order $\epsilon$. We illustrate the deformation procedure by recovering Zakharov's system (\ref{dispersive-Zakharov}) and its Lax pair (\ref{dispersive-Zakharov-Lax}) from their dispersionless limits. To this end, dispersive corrections in the form of infinite series are added to equations (\ref{one-component-dZak}) namely
\begin{equation}
\label{one-component-Zak}
\begin{split}
&u = R, \quad v = V(R) + \epsilon  v_1(R) R_x  + \epsilon^2 (...) + \ldots, \quad
w = W(R) + \epsilon  w_1(R) R_x + \epsilon^2 (...) + \ldots, \\  
&R_t = (R V' + V) R_x + \epsilon \left(a_1(R) R_{xx} + a_2(R) R_x^2\right) + \epsilon^2(...) + \ldots, \\
&R_y = \frac{1}{R (V')^2} R_x + \epsilon \left(A_1(R) R_{xx} + A_2(R) R_x^2\right) + \epsilon^2(...) + \ldots
\end{split}
\end{equation}
where $W' = R (V')^2$ and the unknown functions $v_1(R), w_1(R),
a_1(R), A_1(R), \ldots$ are to be determined. Again, it is assumed
that the relation $u=R$ remains undeformed through a Miura-type
transformation. Equations (\ref{one-component-Zak}) are then assumed
to form a formal solution of some second-order dispersive system
\begin{equation}
\label{dispersive-sigma-Zakharov}
\begin{split}
&u_t = (uv)_x + \epsilon \left( \sigma_{1} u_{xx} + \ldots + \sigma_{22} u_y w_x \right),\quad 
v_t = v v_x + w_x + \epsilon \left( \sigma_{23} u_{xx} + \ldots + \sigma_{44} u_y w_x \right),
\end{split}
\end{equation}
where the nonlocal relation $w_y = u_x$ is kept undeformed and $\sigma_1,\ldots, \sigma_{44}$ are functions of $u,v,w$ to be determined.
Noting that $V(R)$ is an arbitrary function in (\ref{one-component-Zak}), we now require that equations (\ref{one-component-Zak})
define a solution of (\ref{dispersive-sigma-Zakharov}) for {\it any} function $V(R)$. This requirement together with the compatibility condition $R_{ty} = R_{yt}$ specify {\it uniquely} all unknown functions, up to a rescaling of $\epsilon$, and Zakharov's system (\ref{dispersive-Zakharov}) is obtained. Up to the first order, the formal series solution is given by
\begin{equation*}
\begin{split}
&u = R, \quad v = V(R) - \epsilon  (V')^2 R_x  + O(\epsilon^2), \quad
w = W(R) +  O(\epsilon^2), \\  
&R_t = (R V' + V) R_x + O(\epsilon^2),  \quad R_y = \frac{1}{R (V')^2} R_x + O(\epsilon^2),
\end{split}
\end{equation*}
while higher order terms can be determined algorithmically.
We point out that a sufficient condition for defining completely the dispersive deformation, i.e. such that no function remains arbitrary, is that the dispersionless limit must not be linearly degenerate (see Remark \ref{linear-nondegeneracy}). Moreover, it turns out that all systems in the considered class possess a unique integrable deformation. We do not exclude however the possibility that a given dispersionless system may possess two or more inequivalent dispersive counterparts and our algorithm would lead to all such deformations.
Applying the same procedure to systems (\ref{dZak})-(\ref{de8}), we obtain the following exhaustive list of integrable Davey-Stewartson type systems.
\begin{The} Up to the group of point transformations $u\to u,v\to f(u,v)$ and re-scaling of variables, the following systems constitute the complete list of dispersive deformations of systems  (\ref{dZak})-(\ref{de8}) up to order $\epsilon$.
\begin{eqnarray}
\label{Zak} &&\left\{ \begin{array}{l}
u_t=(uv)_x+\epsilon u_{xx}\\
v_t=vv_x+w_x-\epsilon v_{xx} \end{array}, \right.
\\ \nonumber  \\
\label{e1} &&\left\{ \begin{array}{l}
u_t=wu_x+v_x+\epsilon u_{xx}\\
v_t=(vw)_x-\epsilon v_{xx}\end{array}, \right.
\\ \nonumber  \\
\label{e2}&&\left\{ \begin{array}{l}
u_t=(uv)_x+wu_x+\epsilon u_{xx}\\
v_t=(vw)_x+vv_x-\epsilon v_{xx}\end{array}, \right.
\\ \nonumber \\
\label{e3}&&\left\{ \begin{array}{l}
u_t=wu_x-\frac{1}{4}h^2vv_y+v_x+\epsilon (u_{xx}+hv_{xy})\\
v_t=(vw)_x+hvv_x-\epsilon v_{xx}\end{array}, \right.
\\ \nonumber \\ \label{e4}
&&\left\{ \begin{array}{l}
u_t=-wu_y+uu_x+A(v)+\epsilon \left(A^2(u)+u_yA(v)\right)\\
v_t=v^2u_y+2vu_x-wv_y+uv_x+w_x-\epsilon \left(A^2(v)-v_yA(v)\right)\end{array}, \right.
\\ \nonumber \\ \label{e5}
&&\left\{ \begin{array}{l}
u_t=(v-hu)u_x+h(w-\frac{1}{2}v^2)u_y+uv_x-huvv_y+\epsilon\left(B^2(u)-hu_yB(v)\right),\\
v_t=-2hvu_x+h^2v^2u_y+(v-hu)v_x+h(w-\frac{1}{2}v^2)v_y+w_x-\epsilon\left(B^2(v)+hv_yB(v)\right)\end{array} \right.
\\ \nonumber \\
\label{e6}
&&\left\{ \begin{array}{l}
u_t=(hu+v)u_x+(hw-2h^2uv-\frac{1}{2}hv^2)u_y+uv_x-hu(hu+v)v_y\\ \quad\quad\quad+\epsilon\big(B^2(u)-2huB(v_y) -2hu_xv_y-hu_yv_x+3h^2vu_yv_y+2h^2uv_y^2\big),
\\
v_t=-2hvu_x+h^2v^2u_y+(v-hu)v_x+h(w-\frac{1}{2}v^2)v_y+w_x-\epsilon\left(B^2(v)+hv_yB(v)\right)\end{array} \right.
\\ \nonumber \\ \label{e7}
&&\left\{ \begin{array}{l}
u_t=(uv)_x+wu_x-\frac{h}{2}v^2(1+2hu)u_y-huv(1+hu)v_y\\ \quad\quad\quad+\epsilon\big(B^2(u)-2huB(v_y) -2hu_xv_y-hu_yv_x+3h^2vu_yv_y+2h^2uv_y^2\big),
\\
v_t=(vw)_x+vv_x-2hv^2u_x+h^2v^3u_y-\frac{1}{2}hv^2(1-2hu)v_y-2huvv_x-\epsilon\left(B^2(v)+hv_yB(v)\right)\end{array} \right.
\\ \nonumber \\ \label{e8}
&&\left\{ \begin{array}{l}
u_t=(w-2huv)u_x+h^2uv^2u_y+(1-hu^2)v_x+hv(hu^2-1)v_y+\epsilon\left(B^2(u)-hu_yB(v)\right),\\
v_t=-2hv^2u_x+h^2v^3u_y+(w-2huv)v_x+h^2uv^2v_y+vw_x-\epsilon\left(B^2(v)+hv_yB(v)\right)\end{array} \right.
\end{eqnarray}
where $w_y=u_x$ and $A=D_x+vD_y, B=D_x-hvD_y$. Parameter $h$ in the above list if non-zero may be rescaled to 1.
\end{The}

Note that several of these systems have already appeared in other classification works. Zakharov's system (\ref{Zak}) was recovered in \cite{ShabYam1997} along with systems (\ref{e1}) and (\ref{e2}). Moreover the systems
\begin{equation}
u_t = 2 w u_x + u_{xx}, \quad v_t = 2 w v_x - v_{xx}, \quad w_y = (u+v)_x,
\end{equation}
and
\begin{equation}
u_t = 2 w u_x + u_{xx}, \quad v_t = 2 w v_x - v_{xx}, \quad w_y = (uv)_x,
\end{equation}
also appeared in \cite{MikYam1997} and were shown to be equivalent to (\ref{e1}) and (\ref{e2}) through a Miura transformation in \cite{ShabYam1997}. To the best of our knowledge, systems (\ref{e3})-(\ref{e8}) never appeared in other classification problems and seem to be new.

\subsection{Deformation of dispersionless Lax pairs}

Lax pairs for systems (\ref{Zak})-(\ref{e8}) are obtained simultaneously through an analogous procedure.

The dispersionless Lax pairs for systems (\ref{disples-sys}) are given by a pair of equations
\begin{equation}
\label{lc}
F(p,q)=0,\quad s=G(p,q)
\end{equation}
and by relations
\begin{equation}
\label{lc1} p_y=q_x,\quad p_t=s_x,\quad q_t=s_y.
\end{equation}
We can view the dispersionless system (\ref{disples-sys}) together with (\ref{lc}-\ref{lc1}) as a quasilinear system of hydrodynamic type of the form (\ref{displess-system}) with a {\it reducible} dispersion relation (\ref{dispersion-relation}). This system possesses an infinite class of solutions of the form
$$
u=u(R^1,\ldots,R^N),\quad v=v(R^1,\ldots,R^N),\quad w=w(R^1,\ldots,R^N)
$$
$$
p=p(R^1,\ldots,R^N),\quad q=q(R^1,\ldots,R^N),\quad s=s(R^1,\ldots,R^N)
$$
$$
R^i_y=\mu^i(R)R^i_x,\quad R^i_t=\lambda^i(R)R^i_x,\quad i=1,\ldots,N.
$$
Similarly to the construction of deformations for the system  (\ref{disples-sys}) and its hydrodynamic reductions we seek the deformation of the dispersionless Lax pair as
\begin{equation}
\label{pqsdef}
\begin{split}
&p=p(R^1,\ldots,R^N)+\epsilon()+\epsilon^2()+\cdots,\,\, q=q(R^1,\ldots,R^N)+\epsilon()+\epsilon^2()+\cdots,\\ &s=s(R^1,\ldots,R^N)+\epsilon()+\epsilon^2()+\cdots,
\end{split}
\end{equation}
where at $\epsilon^k$ we have homogeneous differential polynomials in derivatives of $R^i$ of degree $k+1$. Deformations of equations (\ref{lc}) are in the form
\begin{equation}
\label{fgdef}\begin{array}{l}
F(p,q)+\epsilon(f_1p_x+f_2p_y+f_3q_y+f_4u_x+f_5u_y+f_6v_x+f_7v_y)=0,\\
s=G(p,q)+\epsilon(g_1p_x+g_2p_y+g_3q_y+g_4u_x+g_5u_y+g_6v_x+g_7v_y+g_8w_x),
\end{array}
\end{equation}
where $f_i=f_i(u,v,p,q)$, $g_i=g_i(u,v,w,p,q)$ and relations (\ref{lc1}) are kept undeformed.
Notice that we need to deform  equations (\ref{lc}) up to the order $\epsilon$ only.

Dispersive deformations of the system (\ref{disples-sys}) are already constructed and thus it is only necessary to find series (\ref{pqsdef}) satisfying (\ref{lc1}) and (\ref{fgdef}).
Analogously to the procedure of constructing the dispersive corrections for the equation one can algorithmically construct series (\ref{pqsdef}) and obtain a set of differential equations on the dispersive deformations $f_i,g_i$.

Equations (\ref{fgdef}), (\ref{lc1}) can be viewed as a nonlinear form of a Lax representation for corresponding dispersive system. To obtain the linear Lax representation we recall that $p=S_x,q=S_y,s=S_t,\psi=e^{\frac{S}{\epsilon}}$ and thus we can apply the transformation
\begin{equation}
\label{lint}
p=\epsilon D_x\log(\psi),\quad q=\epsilon D_y\log(\psi),\quad s=\epsilon D_t\log(\psi),
\end{equation}
to relations (\ref{fgdef}).

Finally, the algorithm of deformations of dispersionless Lax pairs is the following:
\begin{itemize}
\item For a given dispersionless system one obtains its dispersive deformations;
\item One deforms the corresponding dispersionless Lax pairs as in  (\ref{pqsdef}), (\ref{fgdef}) using the dispersive deformation of the system and its hydrodynamic reductions to obtain the nonlinear Lax representation;
\item One applies the linearization transformation.
\end{itemize}

In the following Tables \ref{dispersive-nonlinear-lax-pairs} and \ref{dispersive-linear-lax-pairs} we present nonlinear and linear Lax representations for systems (\ref{Zak})-(\ref{e8}). The transformation between the tables is of the form (\ref{lint}).

We note that for a given dispersive system the deformation of the corresponding dispersionless Lax pair may be not unique.
Moreover for systems  (\ref{Zak})-(\ref{e8}) there exists 3  different dispersive deformations of their Lax pairs.
In the tables we present the simplest deformations only as we need only one Lax representation for every system
(\ref{Zak})-(\ref{e8}) .

\begin{table}[h]
\begin{math}
\begin{array}{|c|l|}
\hline \text{Eq no} & $Nonlinear pair$ \\\hline
(\ref{Zak}) & \begin{array}{l}pq+\frac{1}{2}vq+\frac{1}{4}u+\epsilon
q_x=0,\\ s+p^2+\frac{1}{2}w+\epsilon p_x=0\end{array}  \\ \hline

(\ref{e1}) & \begin{array}{l}pq-\frac{1}{2}up+\frac{1}{4}v+\epsilon
q_x=0,\\ s+p^2-wp+\epsilon p_x=0\end{array}
  \\[1ex] \hline

(\ref{e2})&\begin{array}{l} pq-\frac{1}{2}up+\frac{1}{2}vq+\epsilon q_x=0,\\s+p^2-wp+\epsilon p_x=0\end{array}\\ \hline

(\ref{e3}) &
\begin{array}{l} pq+\frac{1}{4}h v
q+\frac{1}{4}v-\frac{1}{2}up-\frac{1}{8}huv+\epsilon
(q_x+\frac{1}{4}hv_y)=0,\\ s+p^2-wp+\frac{1}{16}h^2v^2+\epsilon
p_x=0
\end{array} \\ \hline

(\ref{e4})&
\begin{array}{l} pq+\frac{1}{2}u q+\frac{1}{4}+vq^2+\epsilon
(vq_y+q_x)=0,\\ s+p^2-v^2q^2+wq+\epsilon (p_x-v^2q_y)=0 \end{array}
\\ \hline

(\ref{e5}) & \begin{array}{l} hpq-\frac{h}{2}(v+hu)q-h^2vq^2+\frac{1}{2}p+\epsilon h(q_x-hvq_y)=0,\\
s+p^2-h^2v^2q^2-\frac{1}{2}h(2w+v^2)q+\epsilon(p_x-h^2v^2q_y)=0\end{array} \\ \hline

(\ref{e6}) & \begin{array}{l}pq-\frac{1}{2}vq-h v q^2+ \frac{1}{4}u+\epsilon(q_x - h v q_y- \left(\frac{1}{2}+h q\right) v_y)=0,\\
s-p^2+h^2v^2q^2 + h ( h uv - hw + \frac{h}{2} v )q - \frac{1}{2}w+\epsilon(h^2 v^2 (q_y - p_x) + (1+2hq)h^2 v^2 v_x )=0\end{array}

  \\ \hline

(\ref{e7}) &
\begin{array}{l}
p q+\frac{p}{2h}(1-hu) - h v q^2+\frac{uv}{4} - \frac{1}{2} v q(1-h u) + \epsilon\left(q_x - h v q_y - h \left(\frac{1}{2h}+q\right)v_y\right) = 0,\\
s+p^2-h^2 q^2 v^2+\frac{1}{2}h u v^2+\frac{1}{2} h q(2 h u-1) v^2 - w p + \epsilon\left(p_x - h^2 v^2 q_y - 2h^2\left(\frac{1}{2h}+q\right) v v_y\right) =0
\end{array}
\\ \hline

(\ref{e8}) & \begin{array}{l} pq-hvq^2-\frac{1}{2}up+\frac{1}{4}v+
\epsilon(q_x-hvq_y)=0,\\ s+p^2-h^2v^2q^2-wp+\frac{1}{4}hv^2+\epsilon(p_x-h^2v^2q_y)=0\end{array}    \\ \hline

\end{array}
\end{math}
\caption{Nonlinear pairs for systems (\ref{Zak})-(\ref{e8}).}
\label{dispersive-nonlinear-lax-pairs}
\end{table}

\begin{table}[h]
\begin{math}
\begin{array}{|c|l|}
\hline \text{Eq no} & $Lax pair$ \\\hline
(\ref{Zak}) &
\begin{array}{l} 4\epsilon^2\psi_{xy}+2\epsilon v\psi_{y}+u\psi=0,\\
2\epsilon\psi_t+2\epsilon^2\psi_{xx}+w\psi=0
\end{array}  \\ \hline

(\ref{e1}) &  \begin{array}{l}
4\epsilon^2\psi_{xy}+v\psi-2\epsilon u\psi_x=0,\\ \psi_t+\epsilon\psi_{xx}-w\psi_x=0\end{array}  \\[1ex] \hline

(\ref{e2})& \begin{array}{l} 2\epsilon\psi_{xy}+v\psi_y-u\psi_x=0,\\
\psi_t+\epsilon\psi_{xx}-w\psi_x=0\end{array}\\ \hline

(\ref{e3}) &
\begin{array}{l}8\epsilon^2\psi_{xy}-4\epsilon u \psi_x+2\epsilon h v \psi_y+(2v-huv+2\epsilon hv_y)\psi=0,\\
16\epsilon\psi_t+16\epsilon^2\psi_{xx}-16\epsilon
w\psi_x+h^2v^2\psi=0\end{array} \\ \hline

(\ref{e4})&
\begin{array}{l}
4\epsilon^2\psi_{xy}+4\epsilon^2v\psi_{yy}+2\epsilon
u\psi_y+\psi=0,\\ \psi_t+w\psi_y-\epsilon
v^2\psi_{yy}+\epsilon\psi_{xx}=0\end{array}
\\ \hline

(\ref{e5}) &
\begin{array}{l}2\epsilon h\psi_{xy}-2\epsilon h^2 v\psi_{yy}+\psi_{x}-h(v+h u)\psi_y=0,\\
2\psi_t+2\epsilon\psi_{xx}-h(v^2+2w)\psi_y-2\epsilon
h^2v^2\psi_{yy}=0\end{array} \\ \hline

(\ref{e6}) &
\begin{array}{l}
4\epsilon^2 \psi_{xy} - 4 h \epsilon^2 v \psi_{yy} - 2 \epsilon(v+2\epsilon h v_y) \psi_y  + (u-2\epsilon v_y)\psi =0,\\
2 \epsilon \psi_t - 2 \epsilon^2 \psi_{xx} +2\epsilon^2h^2v^2\psi_{yy}+\epsilon h(4\epsilon v_x+v^2+2huv-2w)\psi_y+(2\epsilon v_x-w)\psi = 0
\end{array}    \\ \hline

(\ref{e7}) &
\begin{array}{l}
4 h \epsilon^2 (\psi_{xy} - h  v \psi_{yy}) - 2h\epsilon (2h\epsilon v_y + v(1-h u))\psi_y - 2\epsilon(hu-1)\psi_x + h(u v-2 \epsilon v_y) \psi = 0,\\
2\epsilon\psi_t - 2h^2 \epsilon^2 v^2 \psi_{yy} -2 \epsilon w \psi_x + 2 \epsilon^2\psi_{xx}+h \epsilon v ( 2h(uv -2\epsilon v_y)-v) \psi_y+ h v( u v - 2  \epsilon  v_y) \psi = 0
\end{array}
 \\ \hline

(\ref{e8}) & \begin{array}{l}4\epsilon^2\psi_{xy}-4\epsilon^2hv\psi_{yy}-
2\epsilon u\psi_x+v\psi=0,\\ 4\epsilon\psi_t+4\epsilon^2\psi_{xx}-4\epsilon^2h^2v^2\psi_{yy}-4\epsilon w\psi_x+hv^2\psi=0\end{array}   \\ \hline

\end{array}
\end{math}
\caption{Lax pairs of  systems (\ref{Zak})-(\ref{e8}).}
\label{dispersive-linear-lax-pairs}
\end{table}

\section{Conclusion}
The main objective of this paper was to present new classification results based on the method of hydrodynamic reductions and their dispersive deformations. The technique was extended to provide an algorithm to compute the Lax pair of a dispersive system from that of its dispersionless limit. The classification of dispersionless systems was based on the requirement that in order to possess an integrable dispersive deformation, a dispersionless system should be generated by a polynomial Lax pair. In this work, we have restricted ourselves to second degree polynomial generators and a complete list of integrable Davey-Stewartson type systems has been obtained in this context. Consideration of higher degree polynomials may also lead to new systems and this task will be undertaken in a future work.

\section*{Acknowledgements}
The authors would like to thank E.V. Ferapontov and A.V. Mikhailov
for illuminating discussions. The research of BH was supported by a
FQRNT postdoctoral fellowship held within the School of Mathematics
at Loughborough University. Hospitality is gratefully acknowledged.

\end{document}